\documentclass[ ]  {aipproc}

\layoutstyle{6x9}

\SetInternalRegister\hbadness{8000} 

\newcommand\doingARLO[2][]{
  \ifx\mmref\undefined #1\else #2\fi
}
  
\begin{document}

\title [Quantum Treatment]{Quantum Treatment of the Multiple
Scattering and Collective Flow in Intensity Interferometry}

\classification{25.75.Gz}
\keywords{Relativistic heavy-ion collisions, Particle correlations}

\author{Cheuk-Yin Wong}
{
   address={Physics Division, Oak Ridge National Laboratory, 
            Oak Ridge, TN 37830\\
            Physics Department, University of Tennessee, 
            Knoxville, TN 37996}
  ,email={wongc@ornl.gov}
}

\copyrightyear  {2001}

\begin{abstract}
We apply the path-integral method to study the multiple scattering and
collective flow in intensity interferometry in high-energy heavy-ion
collisions.  We show that the Glauber model and eikonal approximation
in an earlier quantum treatment are special examples of the more
general path-integral method.  The multiple scattering and collective
flow lead essentially to an initial source at a shifted momentum, with
a multiple collision absorption factor that depends on the pion
absorption cross section and a phase factor that depends on the
deviations of the in-medium particle momenta from their asymptotic
values.
\end{abstract}

\date{\today}

\maketitle

\vspace*{-0.8cm}
\section{Introduction}

The intensity interferometry, as first proposed by Hanbury-Brown-Twiss
(HBT) to measure the angular diameter of a star \cite{Hbt54}, has been
applied to optical coherence \cite{Gla63}, subatomic physics, and
nuclear collisions \cite{Won94}.  In HBT measurements in high-energy
heavy-ion collisions, the initial source particles undergo multiple
scattering and collective flow.  As a result, it is conventionally
assumed that the initial chaotic source evolves into a different
chaotic source distribution at thermal freeze-out which becomes the
distribution measured in intensity interferometry.  This conventional
assumption is recently subject to question.  Because the intensity
interferometry is purely a quantum-mechanical phenomenon, the multiple
scattering and collective flow must be investigated within a
quantum-mechanical framework \cite{Won03,Won04,Zha04}.  Applying the
Glauber multiple scattering theory at high energies and the optical
model at lower energies, we find that multiple scatterings lead
essentially to an initial source distribution with absorption
\cite{Won03,Won04,Zha04}.  Using the Feynman path integral method, we
further find that the collective flow leads to a phase factor that
depends on the deviations of the in-medium particle momenta from their
asymptotic values.  Subsequent work by Kapusta and Li \cite{Kap04}
supports qualitatively some of the earlier results of Ref.\
\cite{Won03}.  Following this suggestion of a quantum treatment,
Cramer and his collaborators \cite{Cra05} later considered the effects
of produced pions in a static optical model potential.  The work of
Cramer $et~al.$ \cite{Cra05} is however incomplete, as the effects of
particle collisions and the collective flow have not been taken into
account.

In this manuscript, we apply the path-integral method for the quantum
treatment of the multiple scattering and collective flow in intensity
interferometry.  We show that the Glauber model and eikonal
approximation in an earlier quantum treatment of the multiple
scattering \cite{Won03,Won04} are special examples of the more general
path-integral method.

\section{Pion Environment and Elastic Scattering}

Before we present a quantum treatment of intensity interferometry, it
is illuminating to investigate the pion environment after a chaotic
pion source is produced in the phase transition of a quark-gluon
plasma.

At the phase transition temperature of $T\sim$180 MeV, the average
pion density is $\sim$0.3 pions/fm$^3$, the pion energy is $\sim$0.5
GeV, and the average C.M. energy, $\langle \sqrt{s_{\pi \pi}} \,
\rangle$, in a $\pi$-$\pi$ collision is $\sim$0.7 GeV.  Thus the pion
source is dense and pions are energetic after their production.  As
$\langle \sqrt{s_{\pi \pi}} \, \rangle $ is only slightly lower than
the $\rho$ mass, a substantial fraction of $\pi$-$\pi$ collisions will
go through the $I$=1, $\rho$ resonance.  The width of the $\rho$
resonance is 150 MeV in free space.  In a thermalized medium at
$T=150$ MeV, the width increases substantially to $\sim$300 MeV
\cite{Urb02}, and the $\rho$ resonance mean lifetime in the medium
becomes $\sim \hbar/(300 {\rm MeV})$ or 0.67 fm/c.  The orbiting time
for the $\rho$ meson is of order $2\pi r$ and $2r\sim 0.77$ fm
\cite{Won02}.  As the $\rho$ meson mean lifetime in the medium
($\sim$0.67 fm/c) is much shorter than its orbiting time ($\sim$2.4
fm/c), it is unlikely for an $I=1$, $\pi$-$\pi$ pair to complete an
orbital revolution before the $\rho$ meson breaks apart into two
pions.  The scattering of the two pions through the intermediary
$\rho$ meson is essentially an elastic scattering with an
energy-dependent amplitude.  Furthermore, the $\pi \pi \to K \bar K$
threshold is $\sim$1 GeV, substantially greater than $\langle
\sqrt{s_{\pi \pi}} \, \rangle$. Hence, chemical reactions of pions in
the medium are essentially completed very shortly after the phase
transition.  From the state of chemical freeze-out to thermal
freeze-out, the scatterings suffered by the pions are predominately
elastic.

\section{Path Integral Method for Intensity Interferometry}

As a pion propagates from the state of chemical freeze-out to thermal
freeze-out, the elastic $\pi$$\pi$ scattering can be described by
short-ranged scalar and vector interactions, $v_{\rm
col}^{(s,v)}(q-q_i)$, where $q$ is the coordinate of the propagating
pion and $q_i$ is the coordinate of a pion in the medium.  The
propagating pion is also subject to a collective flow which can be
described by a long-range density-dependent mean-field scalar and
vector interactions, $V_{\rm mf}^{(s,v)}(q)$, as in similar cases in
the dynamics of the nuclear fluid \cite{Won78}.  The Lagrangian for
the propagating pion is given by
\begin{eqnarray}
L(q,\dot q)=L_{\rm mf}(q,\dot q)+L_{\rm col}(q,\dot q),
\end{eqnarray}
\vspace*{0.3cm}where
\vspace*{-1.2cm}
\begin{eqnarray}
L_{\rm mf} (q,\dot q) = - [ m_\pi + V_{\rm mf}^{(s)}(q) ] 
                       \sqrt{ 1-{\dot {\bf q}}^2 }
+ {\dot {\bf q}} \cdot {\bf V}_{\rm mf}^{(v)}(q) - V_{\rm mf}^{0(v)}(q),
\end{eqnarray}
\vspace*{-0.6cm}
\begin{eqnarray}
L_{\rm col}(q,\dot q)=- V_{\rm col}^{(s)}(q) 
                        \sqrt{1-{\dot {\bf q}}^2}
+ {\dot {\bf q}} \cdot {\bf V}_{\rm col}^{(v)}(q) - V_{\rm col}^{0(v)}(q),
\end{eqnarray}
\vspace*{-0.8cm}
\begin{eqnarray}
\label{col}
V_{\rm col}^{(s,v)}(q)
=\sum_{i}^{N_\pi-1}
{v}_{\rm col}^{(s,v)}(q-q_i).
\end{eqnarray}
From the Lagrangian of the propagating pion, one obtains the pion
three-momentum,
\begin{eqnarray}
{\bf p}=\partial L/\partial {\dot {\bf q}}=
\gamma [ m_\pi+V^{(s)}(q) ] \dot {\bf q} +{\bf V}^{(v)}(q), 
\end{eqnarray}
the pion Hamiltonian, 
\vspace*{-0.3cm}
\begin{eqnarray}
H=p^0= {\bf p}\cdot {\dot {\bf q}}-L
     =\gamma [ m_\pi + V^{(s)}(q) ] + V^{0(v)}(q),
\end{eqnarray} 
and the pion mass-shell condition,
\begin{eqnarray}
[p^0-V^{0(v)}(q)]^2-[{\bf p}^2-{\bf V}^{(v)}(q)]^2 - [m_\pi+V^{(s)}(q)]^2=0,
\end{eqnarray}
where $\gamma=1/\sqrt{1-{\dot {\bf q}}^2}$ and $V^{(s,v)}(q)=V_{\rm
mf}^{(s,v)}(q)+V_{\rm col}^{(s,v)}(q)$.  For a pion produced at $x$
with momentum $\kappa$ to propagate in the pion medium to the thermal
freeze-out point $x_f$ and be detected at the detecting point $x_d$
with momentum $k$, the probability amplitude is \cite{Kle05}
\begin{eqnarray}
K(\kappa x \to k x_d)
=\int {\cal D}q~ e^{ iS(\kappa x \to k x_d;q) },
\end{eqnarray}
where $\int \!\!{\cal D}q ...$ is the sum over all paths $ q $ from $x$
to $x_d$, and the action $S(\kappa x \to k x_d; q) $ is
\begin{eqnarray}
S(\kappa x \to k x_d;q)
=\int L(q,\dot {\bf q}) ~dt 
= - \int_x^{x_d, ({\rm path~}q)} p(q')\cdot dq'.
\end{eqnarray}
We can separate $S(\kappa x \to k x_d; q)$ into different contributions,
\begin{eqnarray}
\label{eq9}
S(\kappa x \to k x_d;q )
=-k\cdot (x_d - x) +\delta_{\rm mf} (\kappa x, k x_d; q)
+\delta_{\rm col}  (\kappa x, k x_d; q),
\end{eqnarray}
\vspace*{-0.6cm}
\begin{eqnarray}
\label{delmf}
\delta_{\rm mf} (\kappa x,  k x_d; q)
=-\int_x^{x_d,({\rm path~}q)} [ p_{\rm mf}(q')-k ] \cdot dq'
=-\int_x^{x_f,({\rm path~}q)} [ p_{\rm mf}(q')-k ] \cdot dq',
\end{eqnarray}
\vspace*{-0.6cm}
\begin{eqnarray}
\label{eq11}
\delta_{\rm col} (\kappa x, k x_d;q)
=-\int_x^{x_d, ({\rm path~}q)} p_{\rm col}(q') \cdot dq'
=-\int_x^{x_f, ({\rm path~}q)} p_{\rm col}(q') \cdot dq',
\end{eqnarray}
\vspace*{-0.6cm}
\begin{eqnarray}
p_{\rm mf} ( q) 
=\biggl ( \gamma [ m_\pi+V_{\rm mf}^{(s)}( q)]+V_{\rm mf}^{0(v)}(q), 
~~\gamma [ m_\pi+V_{\rm mf}( q) ] \dot {\bf q} 
+{\bf V}_{\rm mf}^{(v)}(q)  \biggr ),  
\end{eqnarray}
\vspace*{-0.6cm}
\begin{eqnarray}
\label{eq13}
p_{\rm col}(q)
=\biggl ( \gamma  V_{\rm col}^{(s)} (q)+V_{\rm col}^{0(v)}(q), 
~~\gamma  V_{\rm col}^{(s)}(q) \dot {\bf q} 
+{\bf V}_{\rm col}^{(v)}(q) \biggr ).
\end{eqnarray}
Because of the additivity of the collision potentials in Eq.\
(\ref{col}), the phase shift for multiple collision $\delta_{\rm col}$
is a sum of the phase shifts for individual collisions, similar to the
case of the Glauber wave function in multiple scattering
\cite{Won03,Won06},
\vspace*{-0.3cm}
\begin{eqnarray}
\delta_{\rm col} (\kappa x, k x_d; q)
=\sum_i^{N_\pi-1} \delta_{{\rm col},i} (\kappa x, k x_d;q),
\end{eqnarray}
where $\delta_{{\rm col},i} (\kappa x, k x_d; q)$ is obtained from
Eqs.\ (\ref{eq11}) and (\ref{eq13}) with the potential $v_{\rm
col}(q-q_i)$ in place of the total collision potential $V_{\rm
col}(q)$.  The propagation amplitude is therefore
\begin{eqnarray}
K(\kappa x \to k x_d)
=\! \! \int \!\! {\cal D}q \exp \{  -ik\cdot(x_d -x)
+ i\delta_{\rm mf} (\kappa x, k x_d; q) 
+ i\delta_{\rm col}(\kappa x, k x_d; q) 
\}.
\end{eqnarray}
If one makes the approximation that the dominant contribution to the
path integral comes from the trajectory along the classical path $q_c$
for mean-field motion (which need not be a straight line), then the
amplitude is approximately
\begin{eqnarray}
K(\kappa x \to k x_d)
\approx \exp \{  -ik\cdot(x_d -x)
+ i\delta_{\rm mf} (\kappa x, k x_d; q_c) 
+ i\delta_{\rm col}(\kappa x, k x_d; q_c) 
\}.
\end{eqnarray}
For the propagation of an energetic pion, the phase shifts along a
straight-line trajectory are just those considered in Ref.\
\cite{Won03}.

\section{Two-Pion Correlations}

For a pion with momentum $\kappa_i$ produced at $x_i$ to propagate
to momenta $k_i$ at the detecting point $x_{di}$, the amplitude is
\vspace*{-0.3cm}
\begin{eqnarray}
\Psi(\kappa_i x_i\to k_i x_{di}) = A(\kappa_i x_i)e^{\phi_0(x_i)} 
   K(\kappa_i x_i\to k_i x_{di}),
\end{eqnarray}
where $A(\kappa_i x_i)$ is the production amplitude, and $\phi_0(x_i)$
is a random and fluctuating production phase for the chaotic
source. The probability amplitude for the production of two identical
pions $(\kappa_1, \kappa_2)$ at $(x_1, x_2)$ to be detected
subsequently as $k_1$ at $x_{d1}$ and $k_2$ at $x_{d2}$ is
\vspace*{-0.3cm}
\begin{eqnarray}
\frac{1}{\sqrt{2}}  \biggl \{ 
\Psi_1(\kappa_1 x_1 \to k_1 x_{d1}) \Psi_1(\kappa_2 x_2 \to k_2 x_{d2})
+( x_1 \leftrightarrow  x_2) \biggr \}.
\end{eqnarray}
The probability $P(k_1 k_2)$ for the detection of two pions with
momenta $(k_1, k_2)$ is the absolute square of the sum of the above
amplitudes from all $x_1$ and $x_2$ source points.  Because of the
random and fluctuating phase $\phi_0(x_i)$ for a chaotic source, the
absolute square of the sum of the amplitudes becomes the sum of the
absolute squares of the amplitudes \cite{Won94}.  One obtains $P(k_1
k_2) = P(k_1)P(k_2) [ 1+R(k_1 k_2)]$, where $P(k_i)$ is the
probability of detecting a pion of momentum $k_i$, and
\begin{eqnarray}
\label{KK}
& &\!\!\!\!\!\!\!\!\!\!\!\!\!\!P(k_1)P( k_2) R(k_1 k_2)
= \sum_{x_1 x_2} 
A(\kappa_1 x_1) A(\kappa_2 x_2) A(\kappa_1 x_2) A(\kappa_2 x_1) 
\nonumber\\
& &\times K(\kappa_1 x_1 \to k_1 x_{d1})K(\kappa_2 x_2 \to k_2 x_{d2})
K^*(\kappa_1 x_2 \to k_1 x_{d1})K^*(\kappa_2 x_1 \to k_2 x_{d2}).
\end{eqnarray}
We can evaluate the product $K(\kappa_1 x_1 \to k_1 x_{d1})
K^*(\kappa_2 x_1 \to k_2 x_{d2})$.  It is equal to
\begin{eqnarray}
K(\kappa_1 x_1 \!\!\!\!\!&\to &\!\!\!\!\!    k_1 x_{d1})
K^*(\kappa_2 x_1 \to k_2 x_{d2})
= e^{-ik_1\cdot(x_{d1}-x_1)+ik_2\cdot(x_{d2}-x_2)} 
\nonumber\\
&&\!\!\!\!\times \int {\cal D}q \int {\cal D}q' 
\exp \{ 
 i\delta_{\rm mf}   (\kappa_1 x_1, k_1 x_{d1};q)
-i\delta_{\rm mf}^* (\kappa_2 x_1, k_2 x_{d2};q')
\nonumber\\
& &~~~~~~~~~~~~~~~~~~~~~~~~~~~+i\delta_{\rm col}  (\kappa_1 x_1, k_1 x_{d1};q)
-i\delta_{\rm col}^*(\kappa_2 x_1, k_2 x_{d2};q')
\}.
\end{eqnarray}
The real part of the phase difference $\delta_{\rm col} (\kappa_1 x_1,
k_1 x_{d1};q) -\delta_{\rm col}^*(\kappa_2 x_1, k_2 x_{d2};q')$ is
stationary when $q-q'=0$ and is random and fluctuating when $q-q'\ne
0$.  The sum over $\exp\{i\delta_{\rm col} (\kappa_1 x_1, k_1
x_{d1};q) -\delta_{\rm col}^*(\kappa_2 x_1, k_2 x_{d2};q')\}$ is
approximately zero when $q-q'\ne 0$.  Thus, the phase factor
$\exp\{i\delta_{\rm col} (\kappa_1 x_1, k_1 x_{d1};q) -\delta_{\rm
  col}^*(\kappa_2 x_1, k_2 x_{d2};q')\}$ operationally behaves
approximately as $ \delta(q-q') \exp\{{-2{\cal I}m~ \delta_{\rm
    col}(\kappa_1 x_1, k_1 x_{d1};q')}\}$.  Upon integrating over
${\cal D} q'$, we obtain
\begin{eqnarray}
\!\!\!\!\!& &\!\!\!\!\!\!\!\!\!\!\!\!\!
K(\kappa_1 x_1 \to k_1 x_{d1}) 
K^*(\kappa_2 x_1 \to k_2 x_{d2})
\approx e^{-ik_1\cdot(x_{d1}-x_1)+ik_2\cdot(x_{d2}-x_2)} 
\nonumber\\
& \times& 
\!\!\!\!\!\!\!\!
\int\!\! {\cal D}q \,
\exp \{  i\delta_{\rm mf}   (\kappa_1 x_1, k_1 x_{d1};q) 
        -i\delta_{\rm mf}^* (\kappa_2 x_1, k_2 x_{d2};q) 
-2{\cal I}m~ \delta_{\rm col}(\kappa_1 x_1, k_1 x_{d1};q)
\}.
\nonumber
\end{eqnarray}
In the sum over paths in the above integral, the dominant contribution
comes from the trajectory $q_c$ that minimizes the action difference
of the mean field, $\{i\delta_{\rm mf} (\kappa_1 x_1, k_1 x_{d1};q)
-i\delta_{\rm mf}^* (\kappa_2 x_1, k_2 x_{d2};q) \}$.  We therefore
have
\begin{eqnarray}
\!\!\!\!\!\!\!\!\!& &\!\!\!\!\!\!\!\!\!\!\!\!\!\!\!
K(\kappa_1 x_1 \to k_1 x_{d1}) 
K^*(\kappa_2 x_1 \to k_2 x_{d2})
\approx \exp \{ -ik_1\cdot(x_{d1}-x_1)+ik_2\cdot(x_{d2}-x_2)
\nonumber\\
& & \!\!\!\!\!\!\! 
+ i\delta_{\rm mf}   (\kappa_1 x_1, k_1 x_{d1};q_c) 
- i\delta_{\rm mf}^* (\kappa_2 x_1, k_2 x_{d2};q_c) 
-2{\cal I}m~ \delta_{\rm col}(\kappa_1 x_1, k_1 x_{d1};q_c)
\}.
\end{eqnarray} 
The other factor in Eq.\ (\ref{KK}) can be approximated in a similar
way.  When all the factors are collected, we obtain
\vspace*{-0.3cm}
\begin{eqnarray}
R(k_1,k_2)= \biggl |
\int d^4x e^{i(k_1-k_2)\cdot x +i\phi_{\rm mf}(k_1k_2,x; q_c)
+i\phi_{\rm col}(k_1k_2,x; q_c)}
\rho_{\rm eff}(k_1k_2,x) \biggr |^2  ,
\end{eqnarray}
where
\vspace*{-1.2cm}
\begin{eqnarray}
\rho_{\rm eff}(k_1 k_1,x)
=\frac {\sqrt{f(\kappa_1 x)f(\kappa_2 x)}} {P(k_1)P(k_2)},
\end{eqnarray}
\vspace*{-0.8cm}
\begin{eqnarray}
\label{eq25}
\phi_{\rm mf}(k_1k_2,x; q_c)=\int_{x}^{x_f,({\rm path~}q_c)} \{ [p_{1\, {\rm mf}} (q)-k_1] 
- [p_{2\, {\rm mf}}^*(q)-k_2] \} \cdot dq,
\end{eqnarray}
\vspace*{-0.6cm}
\begin{eqnarray}
\label{phicol}
\phi_{\rm col}(k_1k_2,x; q_c)=
 \delta_{\rm col}(\kappa_1 x,k_1 x_{d1}; q_c)
-\delta_{\rm col}^*(\kappa_2 x,k_2 x_{d2}; q_c)
\approx 2 i{\cal I}m \delta_{\rm col}(\kappa_1 x, k_1 x_{d1}; q_c),
\nonumber
\end{eqnarray}
and $f(\kappa x)$ is the momentum distribution of the initial chaotic
source at $x$ with momentum $\kappa$ that evolves asymptotically to
$k$ \cite{Won03}.  The collective flow (mean-field interaction)
`distorts' the initial momentum $\kappa$ into the final detected
momentum $k$.  The above results from the path-integral method contain
those of Ref.\ \cite{Won03} as special cases.

The multiple scattering phase $\phi_{\rm col}(k_1k_2,x; q_c)$ can be
simplified to be \cite{Won06}
\vspace*{-0.3cm}
\begin{eqnarray}
-2 {\cal I}m ~\delta_{\rm col}(\kappa_1 x, k_1 x_{d1}; q_c)
=-\int_x^{x_f,({\rm path~} q_c)} n_\pi(q')~\sigma_{\rm abs} ~dq',
\end{eqnarray}
\vspace*{-0.3cm}
where $n_\pi(q')$ is the pion density at $(q')$  and $\sigma_{\rm
abs}$ is the pion absorption cross section.

\vspace*{-0.1cm}
\section{Conclusions and Discussions}

In the environment after the phase transition, chemical reactions are
essentially completed, and the collisions between pions are
predominately elastic. The elastic propagation of a pion can be
studied by the Feynman path-integral method.  We describe the
collective flow of the pion by its dynamical motion in a
density-dependent long-range mean-field potential, and its scatterings
with other pions by short-range {$\pi$-$\pi$} interactions. We find
that HBT correlation measurements lead to an effective source
distribution that depends on the initial source distribution at a
shifted momentum, with a multiple scattering absorption factor
$e^{i\phi_{\rm col}}$ and a collective flow phase factor
$e^{i\phi_{\rm mf}}$.

The multiple scattering absorption factor depends on the pion
absorption cross section. A substantial fraction of the {$\pi$-$\pi$}
collisions will go through the $I$=1, $\rho$ resonance.  The width of
the $\rho$ meson increases substantially in the medium \cite{Urb02}
and the $\rho$ meson mean lifetime in the medium is much shorter than
its orbiting time.  The scattering of two pions through the
intermediary $\rho$ meson is essentially an elastic scattering and
does not represent an absorption process in intensity
interferometry. Therefore, if one considers the $\pi^+$-$\pi^+$
correlation, a detected $\pi^+$ can be absorbed in its propagation
through a pion medium only by interacting with a $\pi^-$ in the
$\pi^+\pi^- \to \pi^0 \pi^0$ reaction.  The cross section
$\sigma(\pi^+\pi^- \to \pi^0 \pi^0)$ is equal to
$(8\pi/9k^2)\sin^2[\delta(I=0)-\delta(I=2)]$, where $k$ is the
magnitude of the pion momentum in the $\pi$-$\pi$ C.M. system and
$\delta(I)$ is the $\pi$-$\pi$ phase shift for the state of total
isospin $I$ \cite{Mar76}. One finds an absorption cross section
$\sigma(\pi^+\pi^- \to \pi^0 \pi^0)$ of $\sim$8 mb and a mean
absorption path length $\sim$12 fm in a thermalized pion medium at
$T$=180 MeV.  The mean absorption path length increases as
the temperature decreases.  The degree of $\pi^+$ absorption by the
$\pi^+\pi^- \to \pi^0 \pi^0$ reaction is small.

The collective flow leads to a net phase shift $\phi_{\rm
mf}(k_1k_2,x; q_c)$ in Eq.\ (\ref{eq25}) that depends on the
deviations of the in-medium particle momenta $p_{i\,{\rm mf}}(q)$ from
their asymptotic values $k_i$.  The contributions of the two terms
from $p_{1\,{\rm mf}}$ and $p_{2\,{\rm mf}}$ in Eq.\ (\ref{eq25}) tend
to cancel and give rise only to a small effect, as indicated by
similar hydrodynamical calculations where one evaluates $\phi_{\rm
mf}$ by following pion trajectories \cite{Zha04}. Because of the small
absorption due to multiple scattering and the small net phase shift
due to the collective flow, the effective density measured in HBT
measurements is expected to depend on a source distribution close to
the initial (chemical freeze-out) source distribution, at a shifted
momentum.  We expect that as the initial pion source transverse
dimension is approximately the spatial dimension of the colliding
nuclei, the effective distribution measured by HBT should be nearly
independent of the collision energy. We also expect that in the
initial source prior to the collective flow, the transverse size in
the ``out'' direction should be approximately the same as that in the
``side'' direction.  Hence, $R_{\rm out}/R_{\rm side}$ should be
approximately close to 1. These expectations are consistent with the
gross features of HBT transverse radii in high-energy heavy-ion
collisions.  It will therefore be of great interest to carry out
numerical calculations of the evolution of the produced hadron matter
to study the above effects.

\vspace*{-0.3cm}
\begin{theacknowledgments}
The author thanks Profs. H. Crater, R. J. Glauber, Chi-Sing Lam, and
Weining Zhang for helpful discussions.  This research was supported in
part by the Division of Nuclear Physics, U.S. Department of Energy,
under Contract No. DE-AC05-00OR22725, managed by UT-Battelle, LLC and
by the National Science Foundation under contract NSF-Phy-0244786 at
the University of Tennessee.
\end{theacknowledgments}


\doingARLO[\bibliographystyle{aipproc}]
          {\ifthenelse{\equal{\AIPcitestyleselect}{num}}
             {\bibliographystyle{arlonum}}
             {\bibliographystyle{arlobib}}
          }

\end{document}